\documentclass[a4paper,10pt,oneside]{article}
\usepackage[english]{babel}
\usepackage{graphicx}
\usepackage{wrapfig,epsfig,epsf}

\begin{document}

\title{Temporal behaviour of global perturbations in compressible axisymmetric flows with free boundaries.}

\author{ 
V.V. Zhuravlev$^{1}$  ,  N.I. Shakura$^{1,2}$
} 

\date{ \it \small
\footnote {e-mail: v.jouravlev@gmail.com}
1) Sternberg Astronomical Institute, Universitetskii pr. 13, Moscow, 119992 Russia  
\\
2) Max-Planck-Institut f\"ur Astrophysik, Karl-Schwarzshild-Str. 1, Postfach 1317, D-85741 Garching, Germany
\\
}

\maketitle

\bigskip 
\hspace{1cm} key words: accretion, accretion disks -- hydrodynamics -- instabilities -- turbulence.
\bigskip

\begin{abstract}
The dynamics of small global perturbations in the form of linear combination of a finite number of non-axisymmetric eigenmodes is studied in two-dimensional approximation. The background flow is assumed to be an axisymmetric perfect fluid with the adiabatic index $\gamma=5/3$ rotating with power law angular velocity distribution $\Omega \propto r^{-q}$, $1.5<q<2.0$, confined by free boundaries in the radial direction. The substantial transient growth of acoustic energy of optimized perturbations is discovered. An optimal energy growth $G$ is calculated numerically for a variety of parameters. Its value depends essentially on the perturbation azimuthal wavenumber $m$ and increases for higher values of $m$. The closer the rotation profile to the Keplerian law, the larger growth factors can be obtained but over a longer time. The highest acoustic energy increase found numerically is of order $\sim 10^2$ over $\sim 6$ typical Keplerian periods. Slow neutral eigenmodes with corotation radius beyond the outer boundary mostly contribute to the transient growth. The revealed linear temporal behaviour of perturbations may play an important role in angular momentum transfer in toroidal flows near compact relativistic objects.
\end{abstract}


\section{Introduction}

Hydrodynamical equations allow the existence of laminar axisymmetric toroidal flows with free boundaries in the vicinity of a gravitating body. The fundamental question is whether the accretion is possible if small perturbations of some sort are present in the medium, i.e whether the perturbations will grow and affect the background to allow the angular momentum transfer. Astrophysically, it is interesting to know if such a transition exists in flows with an almost Keplerian rotation. In the last case setting free boundaries leads to the hypersonic character of the basic flow with Mach number $M>>1$.  

The stability of toroidal flows was studied extensively in the 1980s and 1990s by means of the traditional eigenvalue analysis which has been widely used since the well-known investigations of Lord Rayleigh. The growing non-axisymmetric {\it definite frequency modes} were discovered in the barotropic tori with a positive radial gradient of the specific angular momentum. This phenomenon was cold the Papaloizou-Pringle instability in honour of the authors who published their work in 1984. In numerous subsequent papers this sort of instability was studied in two- and three-dimensional approximations by both analytical and numerical methods \cite{b24,b27,b1,b25,b3,b26,b28,b2,b22}. It was found that there are growing surface gravity and sound waves. The general picture of the instability is constructed by these modes which grow due to the mutual coupling as well as by the Landau mechanism, i.e. due to interaction with the background flow \cite{b8,b5}. However, as the rotation profile approaches the Keplerian law the gravity surface waves become damping while the sonic instability ceases because of very small increments \cite{b3,b4}. This fact actually dropped the subject since the near-Keplerian rotation is of the most interest in connection with the problem of angular momentum transport in the accretion disk theory. Besides, the growing linear modes can exist owing to additional degrees of freedom in the flow. For instance, \cite{b39} and \cite{b40} claimed that stratorotational instability emerges in vertically stratified rotationally stable flows in the shearing sheet approximation.

However, instability of axisymmetric flows can be considered from a quite different viewpoint. It is closely associated with the formulation of the {\it initial value problem} for perturbations. Indeed, to examine the stability thoroughly one must show that {\it any} initial perturbation will not ever affect the background flow. However, the absence of growing eigenmodes guarantees only the asymptotic stability and tells us nothing about the behaviour of perturbations at finite time intervals. The generalized approach to the hydrodynamical instability was realized and declared in the most distinctive form among the fluid physicists in the middle 1990s \cite{b9,b10}. It is often called the nonmodal stability theory and concentrates on the initial value problem, partially on searching for optimal initial perturbations that can exhibit large transient growth even in asymptotically stable flows. Yet the role of this linear mechanism is usually emphasized in connection with the transition to turbulence \cite{b32}. According to the so called bypass concept the linear dynamics is responsible for extraction of the perturbation energy from the background. It is supposed that nonlinear processes only redistribute energy between the modes. The general description and some references to hydrodynamical literature on this subject can also be found in \cite{b13} and in application to astrophysical problems in the more recent paper \cite{b14}. 

Mathematically, the evolution of small perturbations is governed by a linear dynamical operator which in general can be normal or non-normal. In the first case eigenmodes are orthogonal and the eigenanalysis gives the complete picture of the temporal evolution of perturbations. Strictly speaking, the largest possible growth of perturbations is determined at any moment by the highest eigenvalue \cite{b21}. In the case of non-normal dynamical operator eigenmodes become non-orthogonal and the substantial temporal growth of perturbations is possible even if there are no growing eigenmodes \cite{b9}. 

The nonmodal approach to {\it local} perturbations was first pointed out also in astrophysical literature in \cite{b11} and \cite{b12}. The local framework allows us to perform the spatial Fourier transform and thus to study directly the temporal behaviour of perturbations. This direction was developed then in a number of papers \cite{b18,b31,b30,b29,b15,b16,b17}. 

Another branch of astrophysical investigations that use nonmodal approach is devoted to global dynamics in turbulent accretion disks. For instance, \cite{b19} considered the dynamics of small stochastic perturbations in an incompressible Keplerian disk and found the resulting steady-state coherent structures enhancing the angular momentum transfer outward. 
It is also worth mentioning the work \cite{b20} which studied the transient growth of axisymmetric perturbations in a thin accretion disk with account of compressibility and vertical structure of the flow. 

Here we concentrate on the global two-dimensional dynamics of the bounded laminar axisymmetric flows which were studied actively 15-20 years ago. 
We do not introduce viscosity since its action is negligible over the first $\sim 10-50$ rotation periods but we include compressibility which is necessary to do since the sound speed is always comparable or less than the shear velocity difference when we are dealing with free boundaries. First we briefly describe the solution of the boundary problem, i.e. the calculation of eigenmodes. This will be followed by the discussion of the optimization algorithm which allows us to determine the specific group of eigenmodes with the strongest transient growth. This is actually the standard mathematical procedure that has been used by many authors. Then we present the results and make the conclusions.

\section{The Basic Flow and the Boundary Problem for Perturbations.}
We neglect here the vertical structure of the flow and consider an idealized barotropic cylindrical configuration rotating with the angular velocity power profile $\Omega=\Omega_0 r^{-q}$ (here $\Omega_0$ is the Keplerian frequency at $r_0$ where the pressure attains its maximum). To make strict conclusions about the dynamics of real toroidal configurations in the vicinity of a gravitating body one must use full three-dimensional approach. In spite of that it is well known that for a true barotropic flow the problem can be solved using two-dimensional equations with a two-dimensional polytropic index. This is proved analytically for $\lambda>>h$, where $\lambda$ is the wavelength of perturbations and $h$ is the vertical scale of the basic flow (see \cite{b36} for details). Moreover, numerical studies have shown that the perturbed flow behaves almost like in two dimensions even for $\lambda \sim h$ \cite{b1,b2}. The authors themselves explained that by the form of the Reynolds power which does not contain the vertical velocity perturbation in the case of true barotropic configurations. 
Since we are interested in the dynamics of an arbitrary linear combination of the flow eigenmodes with exponential temporal evolution $\propto exp(-i\omega t)$, we should start with solving the boundary problem. For the flow considered here this was done first by \cite{b3} and reexamined for growing perturbations in our previous paper \cite{b4}. The spectrum consists of an infinite number of acoustic modes and may be conventionally divided into three parts. The main part is the so called corotation interval, corresponding to the mode corotation radius (i.e. the point where $\omega=m\Omega$) inside the flow. The corotation interval is given by $(m\Omega(r_2),\,m\Omega(r_1))$, where $r_1$ and $r_2$ are the inner and outer boundary points respectively. Owing to modes coupling and the Landau mechanism \cite{b5,b6} this part of the spectrum contains damping and growing modes which have been of the main interest in previous investigations in the framework of the traditional eigenvalue approach.
However, in flows with almost Keplerian rotation acoustic eigenmodes exhibit very slow growth thus weakly affecting the basic flow. The other two parts of the spectrum correspond to the modes with the corotation radius inside the inner boundary or beyond the outer boundary of the flow. As in this case the energy exchange with the background medium is impossible, they are pure neutral. In the present paper we restrict ourselves to study the latter kind of eigen-solutions both with the growing eigenmodes from the corotation interval. This restriction is done because the Lin rule \cite{b7} forbids to calculate damping and neutral eigenmodes inside the corotation interval by integration along the real axis which has been used in our numerical method in the previous paper \cite{b4}.

The eigen- or, alternatively, definite frequency modes have the standard form: 
$\delta h(r,\varphi,t) = \bar h(r) e^{-i(\omega t - m\varphi)}$. Here $m=1,2,3,...$ is the azimuthal wavenumber and $\omega = \omega_r + i \omega_i$ is generally a complex frequency with real part (also called the pattern speed) and imaginary part (also called the increment). Components of the perturbations velocity field $(\delta v_r,\, \delta v_\varphi)$  have the same form and it is easy to obtain the relations between $\bar v_r,\, \bar v_\varphi$ and $\bar h$ which we omit here but will be necessary below. 

The equation for eigenmodes in a barotropic flow can be 
easily written for $\bar h$:

\begin{equation}
\label{EntalpyEq}
\frac{d}{dr} \left ( \frac{r\rho}{D} \, \frac{d\bar h}{dr}  \right ) - 
\left [   \frac{2m}{\omega - m\Omega}\, \frac{d}{dr} \left ( \frac{\Omega \rho}{D} \right )  +
\frac{m^2}{D r}\, \rho  + \frac{\rho r}{a^2} \right ] \, \bar h = 0 \quad
\end{equation}
where

$$
\hspace{-2.0cm}
\kappa^2 = \frac{2\Omega}{r} \frac{d}{dr} \left ( \Omega r^2 \right ) \,\, \mbox{is the epicyclic frequency}
$$
and

$$
\hspace{-5.5cm}
D = \kappa^2 - (\omega - m\Omega)^2 
$$
$\rho,\,a^2$ are the density and squared adiabatic sound speed profiles of the basic flow, respectively, which can be expressed in terms of the enthalpy $h=a^2/(\gamma-1)$. In turn, $h(r)$ has to be found from the basic flow stationarity expression:
$$
\hspace{-5.7cm}
(\Omega^2-\Omega_g^2) r = \frac{dh}{dr},
$$
where $\Omega_g = \sqrt{GM/r^3}$ is the Keplerian frequency.

The boundary condition is:

\begin{equation}
\label{Boundary}
\Delta h\, |_{r_1,r_2} = \Delta p\, |_{r_1,r_2} = 0\,
\end{equation}
where $\Delta h$, $\Delta p$ are Lagrangian perturbations of the enthalpy and pressure.

Expression (\ref{Boundary}) must be rewritten for $\bar h$:
\begin{equation}
\label{Boundary2}
\frac{dh}{dr}\,\frac{d\bar h}{dr} - 
\left ( D + \frac{1}{r}\,\frac{dh}{dr}\,\frac{2m\Omega}{\omega-m\Omega} \right )\, \bar h \quad 
\Biggl{|_{r_1, r_2}} \Biggr .= 0
\end{equation}

It turns out that (\ref{Boundary2}) is equivalent to the regularity condition for the solution at the boundary points, which must be imposed since coefficients of eq. (\ref{EntalpyEq}) become singular at $r_1$ and $r_2$. 

Eq. (\ref{EntalpyEq}) with b.c. (\ref{Boundary2}) was solved for a variety of parameters in \cite{b4}. As an example, the part of the spectrum for $q=1.58,\,w=300\%,\,\gamma=5/3$ and $m=5$ is shown in Fig.~1. In this figure the first several neutral modes with $\omega<m\Omega(r_2)$ and some growing modes from the corotation interval with $\omega_i>10^{-4}$ are displayed. Here and below all frequencies are given in units of $\Omega_0$ and time in units of $\Omega_0^{-1}$; we also introduce $w=(r_2-r_1)/r_0 \times 100\%$, the radial size of the flow in percents. We also use the Mach number defined as
$M=(\Omega(r_1)r_1 - \Omega(r_2)r_2)/a(r_0)$ though the presence of free boundaries where $p=a=0$ and non-trivial distribution of $a(r)$ makes $M$ to be not so obvious parameter to describe the compressibility of the flow. In real three-dimensional flows (i.e. tori) around a gravitating body $M$ approximately equals to $2(r_2-r_1)/h$, which is the ratio of the torus radial size to its half-thickness.

\begin{figure}
\epsfxsize=12cm
\centerline{\epsfbox{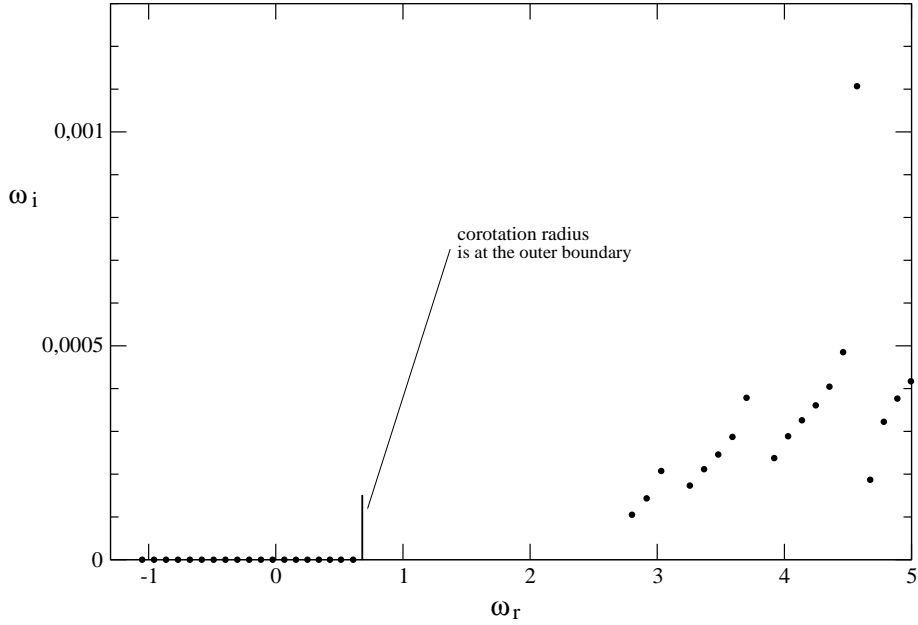}}
\caption
{The part of the spectrum for acoustic modes. The set of neutral eigenmodes outside of the corotation interval with $\omega<m\Omega(r_2)$ and some growing eigenmodes with $\omega_i>10^{-4}\Omega_0$ and $\omega_r<m\Omega_0$ can be seen. $q=1.58$, $w=300\%$, $\gamma=5/3$, $m=5$.} 
\end{figure}

\section{The Optimization Problem}

\subsection{Class of perturbations and the growth measure choice}

Let us consider a linear subspace of the solutions of the perturbation equations, which is represented by the linear span $S_N$ of a finite number of eigenmodes:

\begin{equation}
\label{lin_comb}
q = \sum_{n=1}^{N} \kappa_n \tilde q_n
\end{equation}

The system $\{ \tilde q_n \}$ forms the basis in this linear subspace and each basis vector has the form:
$$
\hspace{-1.5cm}
\tilde q_n = \{ \bar v_r,\, \bar v_\varphi,\, \bar h \}_n \times e^{i m \varphi} = \{\tilde q^1,\,\tilde q^2, \, \tilde q^3 \}_n \times e^{i m \varphi}, 
$$
where $\{ \bar v_r, \bar v_\varphi, \bar h \}_n$ are solutions to the one-dimensional boundary problem described in the previous section.
We assume that the coordinates of $q$ in $S_N$ reflect the temporal behaviour of $q$:
$$
\hspace{-5.8cm}
\kappa_n (t) = \kappa_n^0 e^{-i \omega_n t},
$$
where $\kappa_n^0$ are arbitrary complex numbers and $\omega_n$ are eigen-frequencies of the corresponding eigenmodes $\tilde q_n$. 

Thus, the temporal evolution of $q$ is dictated by the matrix expression for its coordinates:
\begin{equation}
 \label{evolution}
\mbox{\boldmath$\kappa$} = e^{\mbox{\boldmath$\Lambda$} t} \mbox{\boldmath$\kappa$}^0,
\end{equation}
where $\mbox{\boldmath$\kappa$}^0 = (\kappa_1^0,\kappa_2^0,...,\kappa_N^0)^T$, 
$\mbox{\boldmath$\kappa$} = (\kappa_1,\kappa_2,...,\kappa_N)^T$ and the exponential is the propagator acting on the initial perturbation; evidently, $\mbox{\boldmath$\Lambda$} = diag\{-i\omega_1,-i\omega_2,...,-i\omega_N\}$.

The next fundamental step is to introduce the inner product to measure the specific perturbation magnitude and its evolution in time. This is done such that the norm equals to the total acoustic energy of the perturbation:
\begin{equation}
 \label{En}
E_a = \frac{1}{2} \int \, \rho \left ( \delta v_r^2 + \delta v_\varphi^2 + \frac{\delta h^2}{a^2} \right ) \, rdrd\varphi
\end{equation}
In a bounded flow the latter satisfies the equation \cite{b3}:
\begin{equation}
 \label{en_eq}
\frac{d E_a}{d t} = -\int r \frac{d \Omega}{d r} \rho \delta v_r \delta v_\varphi
r\, dr\, d\varphi  
\end{equation}
Sound waves cannot carry energy out of the fluid so the integral of the acoustic energy flux across the flow boundary vanishes and $E_a$ changes only due to the work done by the Reynolds forces. 
Consequently, it seems natural to define the perturbation norm via (\ref{En}), since the evolution of $E_a(t)$ reflects how much energy a given perturbation obtains from the basic flow.
Next,
$$
\hspace{-1.5cm}
E_a = \frac{1}{2} \int \rho \left [ \left(Re(q^1 e^{im\varphi})\right)^2 + \left(Re(q^2 e^{im\varphi})\right)^2 + \right .
$$
$$
\hspace{-4.2cm}
\left . + \; \frac{1}{a^2} \left(Re(q^3 e^{im\varphi})\right)^2 \right ] rdrd\varphi = 
$$
$$
\hspace{-2.7cm}
=\pi \int\limits_{r_1}^{r_2} \rho \left [ {q^1}^* q^1 + {q^2}^* q^2 + \frac{1}{a^2} {q^3}^* q^3 \right ] r dr,
$$
where the asterisk denotes the complex conjugation.

Hence, the inner product between two arbitrary vectors of the linear span $(f,g)$ must have the form:
\begin{equation}
\label{inn_prod}
(f,g) = \pi \int\limits_{r_1}^{r_2} \rho \left [ {f^1}^* g^1 + {f^2}^* g^2 + \frac{1}{a^2} {f^3}^* g^3 \right ] r dr
\end{equation}
Finally, the vector norm, i.e. the total perturbation energy is
\begin{equation}
 \label{norm}
(q,q) = E_a = \|\mbox{\boldmath$\kappa$}\|^2 = \sum_{i,j=1}^{N} \kappa_i^* \kappa_j M_{ij} = \mbox{\boldmath$\kappa$}^\dag {\bf M} \mbox{\boldmath$\kappa$}
\end{equation}
with $M_{ij} = (\tilde q_i,\, \tilde q_j)$
being the metrics in $S_N$; Here $\dag$ stands for the Hermitian transpose.

To conclude this section, we define the growth factor as:
\begin{equation}
\label{g_t}
g(t) = \frac{\| \mbox{\boldmath$\kappa$}(t) \|^2}{\| \mbox{\boldmath$\kappa$}^0 \|^2} = 
\frac{\| e^{\mbox{\boldmath$\Lambda$}t} \mbox{\boldmath$\kappa$}^0 \|^2}{\| \mbox{\boldmath$\kappa$}^0 \|^2}
\end{equation}

\subsection{The optimal growth}

An instability study looks for an initial perturbation with maximum growth in a finite time interval. This is a kind of variational problem when one should fix the time interval $T$ and find the most suitable $\mbox{\boldmath$\kappa$}^0$ giving the highest value of $g(T)$. The mathematical formulation of this problem has been in common use in the hydrodynamical literature \cite{b38,b35,b37}. 

We need to calculate the propagator norm (see eq. (\ref{evolution})~), since by definition it is equivalent to the optimal growth $G$ (cf. (\ref{g_t}) ):
\begin{equation}
\label{G_T}
G(T) = \max_{\mbox{\boldmath$\kappa$}^0\neq 0} g(T) \equiv \| e^{\mbox{\boldmath$\Lambda$} T} \|^2
\end{equation}
The matrix ${\bf M}$ is positive definite, thus the Cholesky decomposition is permissible: ${\bf M} ={\bf F}^\dag {\bf F}$. This means that we can transit to the Euclidean metrics in $S_N$:
$\| \mbox{\boldmath$\kappa$} \|^2 = \| {\bf F}\mbox{\boldmath$\kappa$} \|_2^2$. 
Thus, 
$$
\hspace{-1.8cm}
G(T) = \max_{\mbox{\boldmath$\kappa$}^0\neq 0} \frac{\| e^{\mbox{\boldmath$\Lambda$}t} \mbox{\boldmath$\kappa$}^0 \|^2}{\| \mbox{\boldmath$\kappa$}^0 \|^2} = 
\max_{\mbox{\boldmath$\kappa$}^0\neq 0} \frac{\| {\bf F}e^{\mbox{\boldmath$\Lambda$}t} \mbox{\boldmath$\kappa$}^0 \|_2^2}{\| {\bf F}\mbox{\boldmath$\kappa$}^0 \|_2^2} =
$$
\begin{equation}
\label{G_T_add}
=\max_{\mbox{\boldmath$\kappa$}^0\neq 0} \frac{\| {\bf F}e^{\mbox{\boldmath$\Lambda$}t} {\bf F}^{-1}{\bf F}\mbox{\boldmath$\kappa$}^0 \|_2^2}{\| {\bf F}\mbox{\boldmath$\kappa$}^0 \|_2^2} \equiv \| {\bf F}e^{\mbox{\boldmath$\Lambda$} T} {\bf F}^{-1} \|_2^2
\end{equation}
 The last term in (\ref{G_T_add}) is actually the two-norm of a $N \times N$ complex matrix which is equal to the first (the largest) singular value, consequently:
$G(T)=\sigma_1^2\left({\bf F}e^{\mbox{\boldmath$\Lambda$}T} {\bf F}^{-1} \right)$. 

The singular value decomposition of ${\bf F}e^{\mbox{\boldmath$\Lambda$}T} {\bf F}^{-1}$ 
gives $\sigma_1$ and the associated right singular vector ${\bf v}_1$. The initial unit vector $\mbox{\boldmath$\kappa$}^0 = {\bf F}^{-1} {\bf v}_1$ 
corresponds to the perturbation that attains the largest possible energy growth $G$ at the moment $T$.  

\vspace{0.5cm}
{ \noindent \bf Comment on the numerical method}
\vspace{0.2cm}

We have seen that the optimization procedure reduces to a few numerical steps. First, we choose a finite set of eigenvalues $\omega_n$ and precisely calculate the radial profiles of eigenmodes $\{ \bar v_r, \bar v_\varphi, \bar h \}_n$. The metrics ${\bf M}$ is determined then by the numerical integration according to (\ref{inn_prod}). This is the most time consuming task. When we obtain ${\bf M}$ the standard linear algebra methods allow us to get $G(t)$ as well as $g(t)$ for any perturbation with the given initial condition $\mbox{\boldmath$\kappa$}^0$.

\section{Results}
\subsection{Combinations of eigenmodes}

\begin{figure}
\epsfxsize=12cm
\centerline{\epsfbox{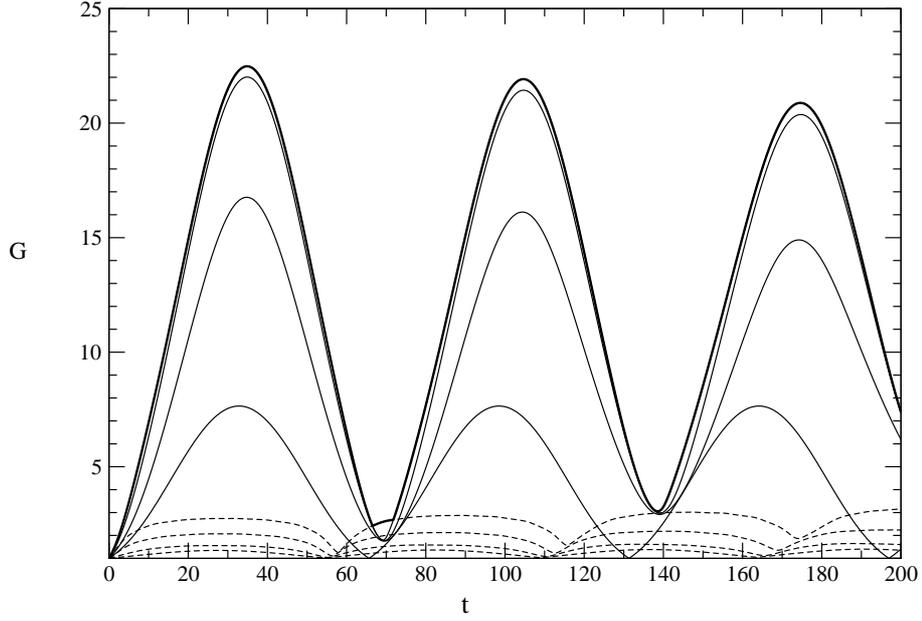}}
\caption
{The optimal transient growth for a linear combination of different sets of eigenmodes. From bottom to up: 
dashed curves show 2,4,10,19 growing eigenmodes, solid curves show 2,4,10 neutral eigenmodes (displayed in Fig.~1). The bold curve represents the optimal transient growth for a linear combination of both 19 neutral and 19 growing eigenmodes and covers the solid curve that corresponds to 19 neutral eigenmodes. The fixed parameters are the same as in Fig.~1.
 }
\end{figure}

\begin{figure}
\epsfxsize=12cm
\centerline{\epsfbox{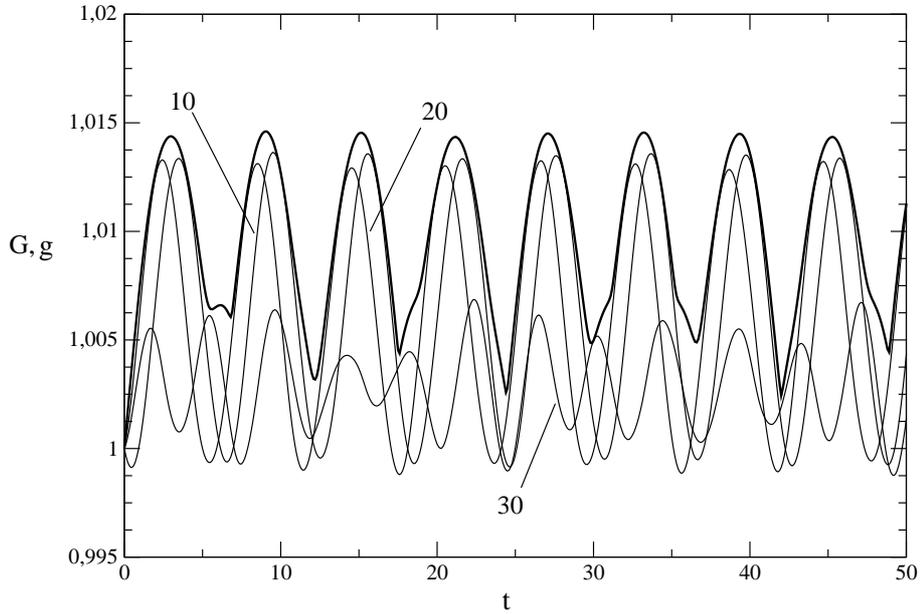}}
\caption
{The bold curve is the optimal transient growth for a linear combination of the first 5 neutral eigenmodes with $\omega>m\Omega(r_1)$. Other curves denote the energy behaviour of specific initial perturbations depending on time. These are the perturbations that attain the largest possible growth at the moments $t$ marked by corresponding numbers. The fixed parameters are the same as in Fig.~1.
 }
\end{figure}

First we would like to clarify which combinations of eigenmodes can achieve the most substantial energy growth. We take the part of the spectrum displayed in Fig.~1 and calculate 
$G(t)$ separately for groups of 2,4,10,19 neutral modes when the corotation radius is beyond the outer boundary and for groups of 2,4,10,19 growing modes with $\omega_i>10^{-4}$ and $\omega_r<m\Omega_0$ (see Fig.~2). Obviously, a number of neutral modes is able to exhibit a much greater energy increase rather than the growing modes. Moreover, the unification of all 38 modes does not affect $G$ noticeably. This is illustrated by the bold curve. One can also see that $G$-curve is slightly modified when passing from 10 to 19 neutral modes. The numerical test shows that this is the case for larger combinations of neutral modes. Furthermore, various combinations of the growing modes with $\omega_r>m\Omega_0$, i.e. with the corotation in the inner part of the flow, were examined and did not reveal any substantial growth. 

The fact that slow neutral modes are of the most interest from the standpoint of transient growth is also supported by the calculation of $G$ for the fast neutral modes with $\omega>m\Omega(r_1)$, i.e. with the corotation radius inside the inner boundary of the flow. The result is shown in Fig.~3. Here the curves $g(t)$ are also added to see the temporal behaviour of specific initial perturbations. Mathematically, the weak energy growth is explained by small off-diagonal elements in ${\bf M}$ and, consequently, by an almost normal propagator that governs the initial disturbances. 

If the propagator is fully normal $G=g=const=1$.
In fact, for the unit $\mbox{\boldmath$\kappa$}^0$ eq. (\ref{g_t}) gives:
$$
\hspace{-1cm}
g(t) = 
\| e^{\mbox{\boldmath$\Lambda$}t} \mbox{\boldmath$\kappa$}^0 \|^2=
\sum_{n=1,k=n}^N {\kappa_n^o}^* e^{i\omega_n^* t}\, \kappa_k^o\, e^{-i\omega_k t} M_{nk} =
$$
\begin{equation}
\label{normal_case}
=\sum_{n=1}^N |\kappa_n^o|^2\, e^{2\omega_i^n t} M_{nn} 
\end{equation}
which is unity for neutral modes and is bounded by the largest increment for growing modes. 

Thus we conclude that the optimal growth arises mostly when considering the first 10-15 slow neutral modes with $\omega<m\Omega(r_2)$. 
We keep it in mind and present below the results concerning this kind of eigenmodes.

Besides, one can notice that there is the typical scale of $G$ variation. Though $G$ starts from unity, it never returns back (see Fig.~2) and has extrema separated by the characteristic time interval which is evidently determined by the shortest difference $\omega_i-\omega_j$ that is present in ${\bf F}e^{\mbox{\boldmath$\Lambda$}T} {\bf F}^{-1}$ and equals to the step between the eigenvalues in the spectrum. Then, the shorter the variation time, the weaker is the transient growth. 
Note that this dependence is very steep (cf. Fig.~2 and Fig.~3).

\subsection{Parametric study of the optimal growth}

In this section we examine the temporal growth by varying free parameters of the problem, i.e. the azimuthal wavenumber $m$, the flow radial size $w$, and parameter $q$ which defines the rotational profile and pressure distribution in the basic flow. The last one is certainly important since it describes the deviation from the pure Keplerian rotation and hypersonic feature of the flow. The research is carried out for the particular case $\gamma=5/3$ specifying the compressibility of the fluid. We restrict ourselves to study the combination of the first 10 slow neutral modes, as discussed above.

Let us first look what happens when we change $m$, i.e. the characteristic length scale of perturbations. The result is shown in Fig.~4. The optimal growth increases with azimuthal wavenumber starting from $m=2$. So perturbations with smaller length scale have more capacity to gain energy from the background. Note that the timescale of $G$ variations weakly depends on $m$ and is nearly $~70\Omega_0^{-1}$ or $~10T_0$, where $T_0$ is the Keplerian period at $r_0$. 
Furthermore, the first maximum of $G$ occurs at $\sim 6T_0$ for the above parameters. The exception is the curve corresponding to $m=1$. It has a longer timescale of variations and attains higher values of $G$.

\begin{figure}
\epsfxsize=12cm
\centerline{\epsfbox{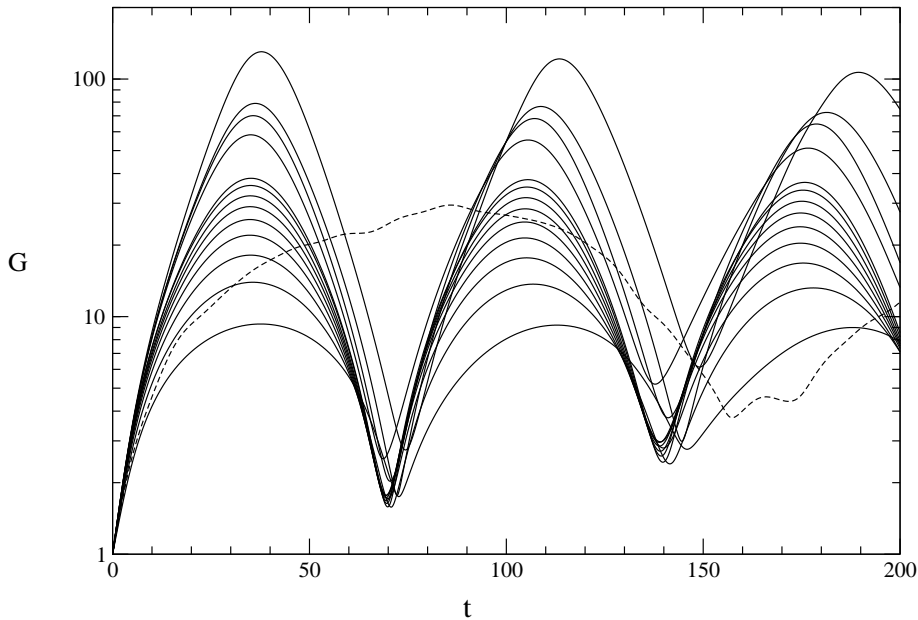}}
\caption
{The optimal transient growth for a linear combination of the first 10 neutral eigenmodes with $\omega<m\Omega(r_2)$ corresponding to different azimuthal wavenumbers. The dashed curve corresponds to $m=1$. For solid curves from bottom to up: $m=2,3,4,5,6,7,8,9,10,15,20,30,50$. $q=1.58$, $w=300\%$, $\gamma=5/3$.
 }
\end{figure}

In Fig.~5 and Fig.~6 we present the optimal growth curves for different $w$ and $q$. The later $G$ attains its maximum value, the greater energy growth of perturbations is possible. As we have noticed before, the large scale variation of $G(t)$ is defined by the step in the acoustic spectrum which decreases while passing to more extended flows or approaching the Keplerian rotation. The last fact also implies the Mach number growth. The explicit dependence on the Mach number $M$ defined in the second section of the paper is shown in Fig.~7. Here $G_{max}$ is the first maximum in the optimal growth curve and $T_{max}$ is the time at which $G=G_{max}$.
Note that there are several steep changes in the overall monotonic growth of $G_{max}$. This is due to sudden changes in the set of eigenvalues and eigenfunctions we take to analyse the optimal growth. Indeed, by convention we take the first 10 neutral modes with the corotation radius beyond the outer boundary, so our frequency frame is $\omega_{cr}=m\Omega(r_2)$. At the same time the acoustic spectrum reveals the continuous shift with respect to $\omega_{cr}$ due to variation of free parameters. Since the spectrum is discrete there are moments when some eigenvalue crosses the point $\omega_{cr}$ and appears (disappears) in the spectral part that we examine.

\begin{figure}
\epsfxsize=12cm
\centerline{\epsfbox{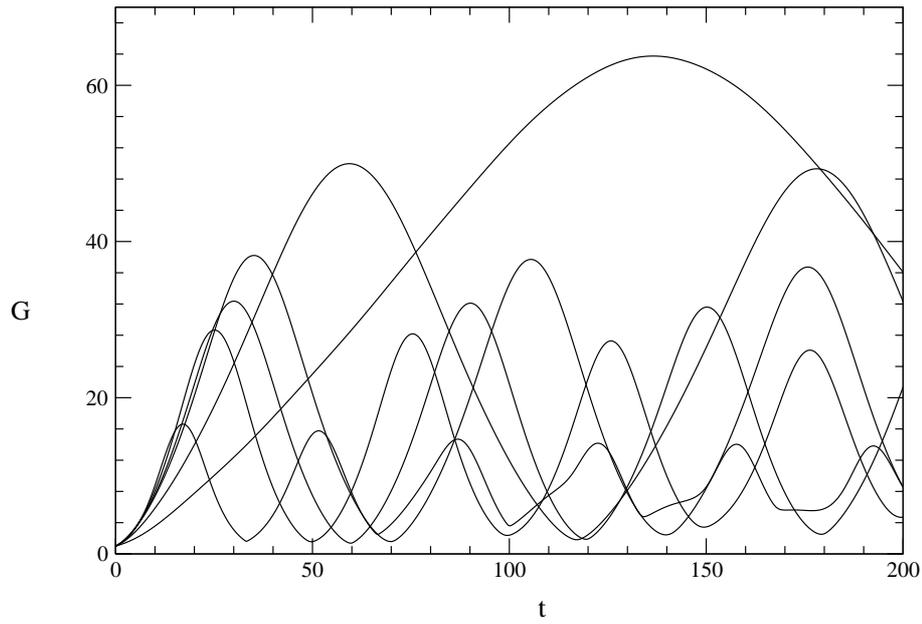}}
\caption
{The same as in Fig.~4 but for $m=10$ and different values of w. For maxima of $G$ (from bottom to up): $w=100\%, 200\%, 250\%, 300\%, 500\%, 1000\%$.
 }
\end{figure}

However, in general view $G_{max}$ and $T_{max}$
are pretty proportional to the Mach number, at least in the range of $q$ studied numerically. 
These dependencies can be approximately written as:
$$
\hspace{-4.7cm}
G_{max} \simeq 38 + 5.6 \, (M-5) 
$$
$$
\hspace{-4.7cm}
T_{max} \simeq 5.5 + 1.2\, (M-5)
$$
Note that here $T_{max}$ is given in units of $T_0$ for conveniency.
Let us remind that for a barotropic flow the Mach number approximately equals to the ratio of the radial size of the torus to its half-thickness: $M \simeq 2(r_2-r_1)/h$. 
We should mention here as well that \cite{b35} and \cite{b34} studied the boundary layer and the plane Couette problem (and an unbounded constant shear) correspondingly. They included both viscosity and compessibility and also found that the capacity to transient growth increases with increasing Mach number although this dependence seems to be less strong than in the present research. In fact, this may be due to influence of viscosity. 

This result is quite important, since in astrophysics we are dealing with almost Keplerian and thus highly supersonic flows. It is remarkable that hypersonic feature of astrophysical flows was indeed the obstacle in the traditional eigenvalue analysis since the well-known acoustic instability vanishes while proceeding to high Mach numbers. In spite of this fact the specially selected {\it combination} of eigenfunctions is able to reach even the larger energy growth for almost Keplerian flows with $q \to 3/2$.

\begin{figure}
\epsfxsize=12cm
\centerline{\epsfbox{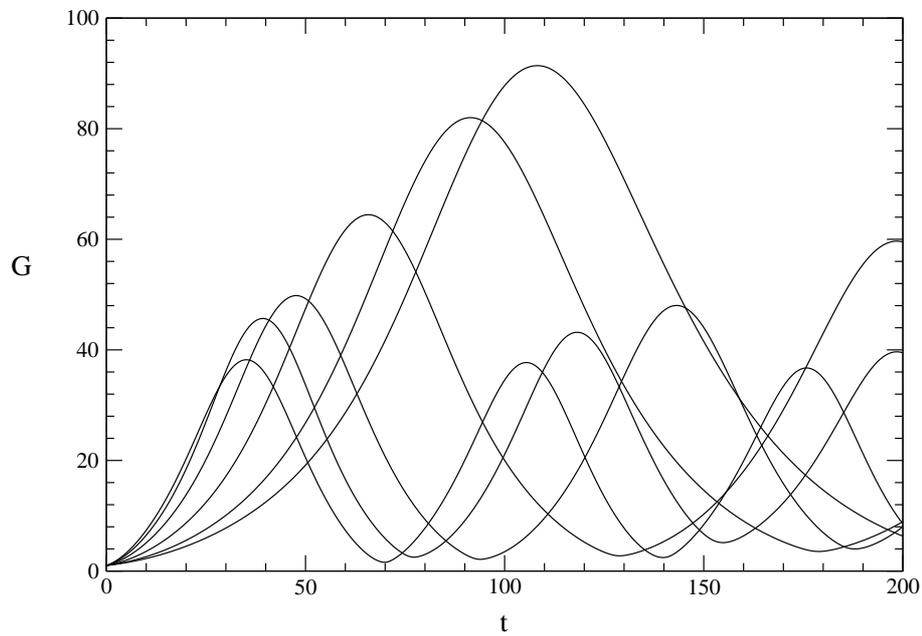}}
\caption
{The same as in Fig.~5 but for $w=300\%$ and different values of $q$. For maxima of $G$ (from bottom to up): $q=1.58, 1.56, 1.54, 1.52, 1.51, 1.507$.
}
\end{figure}

However, one must still keep in mind that the latter conclusion is done in two-dimensional approach and must be checked for group of eigenmodes accurately calculated in three dimensions. Nevertheless, we hope this remains valid since the transient growth fenomenon arises from the main feature of the flow considered, i.e. the shear.  

\begin{figure}
\epsfxsize=12cm
\centerline{\epsfbox{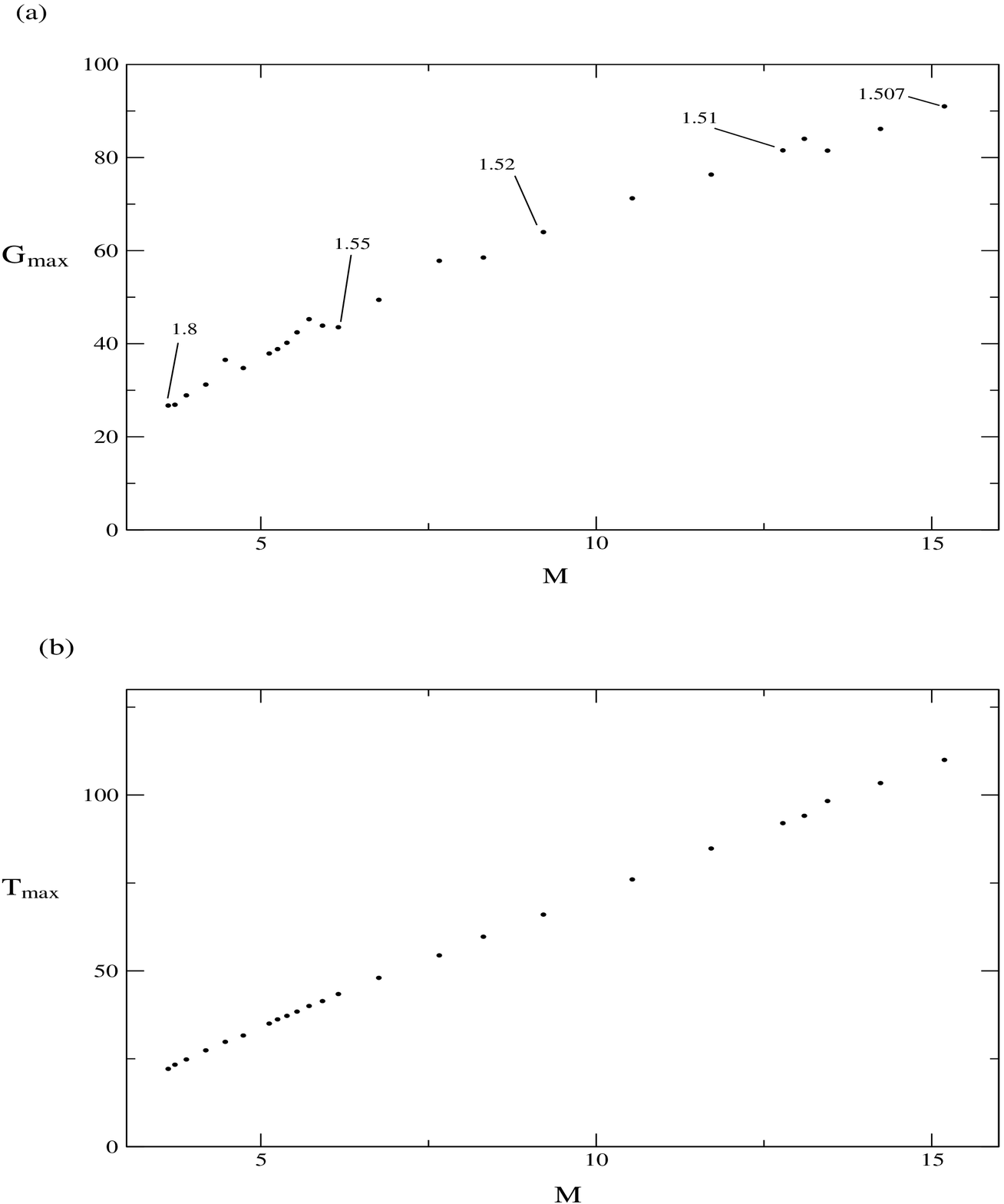}}
\caption
{$G_{max}$ and $T_{max}$ depending on Mach number as defined in the text. The numbers denote some of the corresponding values of $q$.
$w=300\%$, $m=10$, $\gamma=5/3$.}
\hspace{1cm}
\end{figure}

The last two plots are in a sense an addition to the results obtained. Fig.~8 shows the temporal behaviour of individual perturbations which is drawn over by the bold curve that represents the optimal growth. The numbers denote the time $T$ when the corresponding curves touch the bold one, i.e. when the specific perturbations reach the largest possible growth factor $G(T)$. 
At last we present Fig.~9 to show the long time behaviour of the optimal growth. The overall result is that starting from unity $G(t)$ never returns back despite variations with a notable amplitude - the feature that we noticed before when commenting on Fig.~2. Here this is confirmed for a longer time interval. The consequence is that one can always choose initial perturbations that will reveal a substantial growth at any moment in future. Certainly, the necessary condition for this property to exist is the absence of dissipation in our approach.

\begin{figure}
\epsfxsize=12cm
\centerline{\epsfbox{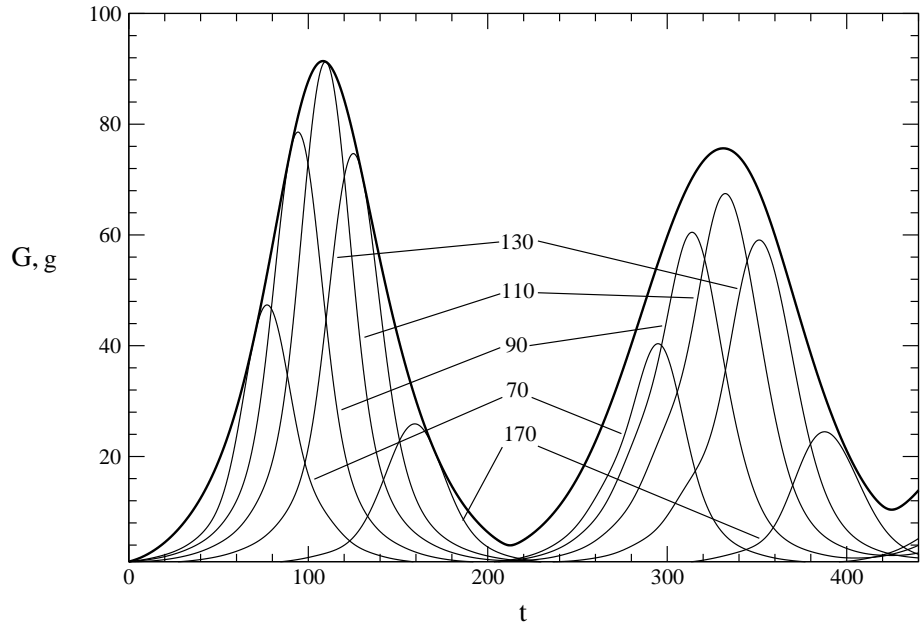}}
\caption
{ The bold curve is the optimal transient growth displayed for $w=300\%$, $\gamma=5/3$, $m=10$ and $q=1.507$ in Fig.~6. The other curves denote the energy behaviour of a specific initial perturbations depending on time. These are the perturbations that attain the largest possible growth at the moments of $t$ marked by the corresponding numbers.
 }
\end{figure}

\begin{figure}
\epsfxsize=12cm
\centerline{\epsfbox{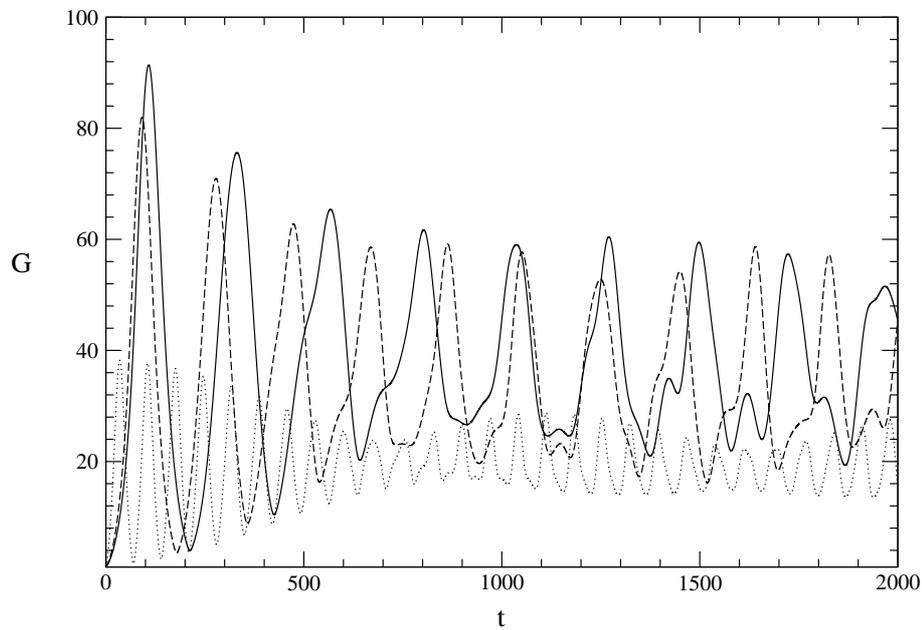}}
\caption
{ The same as in Fig.~6 but in a longer time interval. The dotted, dashed and solid curves have been calculated accordingly for $q=1.58, 1.51, 1.507$.
}
\end{figure}

\newpage

\section{Summary}

Our study shows the significance of the transient growth concept in application to astrophysical problems. We considered here the dynamics of {\it global} perturbations in a sense that free boundary conditions which are natural in astrophysical practice were involved. Previous studies of a pure hydrodynamical instability in axisymetric flows with free boundaries have been carried out by methods of the traditional eigenanalysis, i.e. by searching for growing modes with exponential dependence on time. This way was proved to be unable to solve the problem of angular momentum transfer in almost Keplerian flows with finite compressibility, since the surface gravity eigenmodes are damping \cite{b3} and the growing sonic eigenmodes have negligible increments. However, the present two-dimensional analysis demonstrates that situation noticeably changes if one turns to the dynamics of a linear {\it combination} of eigenmodes. In fact, we studied mainly a finite number of slow neutral eigenmodes from acoustic spectrum of the basic flow with $\omega<m\Omega(r_2)$. It turns out that such a combination exhibits a substantial transient acoustic energy growth $g$ on the dynamical timescale. Its magnitude strongly depends on parameters of the problem, i.e. the azimuthal wavenumber of the modes, the radial size and the rotation profile of the background flow. The numerical tests produce $g\propto 10^2$ in a few typical Keplerian periods. Possibly the most notable fact is that $g$ grows as we approach the Keplerian rotation. This result is opposite to the conclusions of eigenanalysis according to which the sonic instability ceases to affect the basic flow just because of the small sound speed. If one introduces the Mach number $M$ as the ratio of the shear velocity difference between the boundaries $r_1$ and $r_2$ to the maximum sound speed in the flow, then the maximum optimal growth and the corresponding time interval are approximately proportional to $M$. So the results of the present two-dimensional analysis indicate that the flat hypersonic tori in the vicinity of gravitating objects can exhibit a large linear transient growth of global perturbations which may be related to the problem of angular momentum transfer. However, the conclusions we made here certainly must be checked by complete three-dimensional calculations. 

\vspace{1cm}

{\noindent \bf Acknowledgements}

\vspace{0.2cm}

We would like to thank K.A. Postnov for careful reading of the manuscript.
This paper was supported by grant RFFI 06-02-16025.

\end{document}